\documentstyle[epsfig,aas2pp4]{article}

\lefthead{B. Paul et al.}
\righthead{Long term emission stability of two AXPs}

\newcommand\asca{{\it ASCA}}
\newcommand\exo{{\it EXOSAT}}
\newcommand\xte{{\it RXTE}}
\newcommand\ros{{\it ROSAT}}
\newcommand\gin{{\it GINGA}}
\newcommand\ein{{\it Einstein}}
\newcommand\sax{{\it BeppoSAX}}
\newcommand\axpa{4U~0142+61}
\newcommand\axpb{1E~1048.1$-$5937}
\newcommand\sgra{SGR~1806$-$20}
\newcommand\sgrb{SGR~1900+14}

\begin{document}

\title{Study of the Long Term Stability of two Anomalous X-ray Pulsars
\axpa\ and \axpb\ with \asca}

\author{B. Paul,\footnotemark[1] M. Kawasaki, T. Dotani, and F. Nagase}

\affil{The Institute of Space and Astronautical Science,\\
3-1-1 Yoshinodai, Sagamihara, Kanagawa 229-8510, Japan;\\
bpaul@astro.isas.ac.jp, kawasaki@astro.isas.ac.jp,
dotani@astro.isas.ac.jp, nagase@astro.isas.ac.jp}
\footnotetext[1]{On leave from the Tata Institute of Fundamental
Research, Homi Bhabha road, Mumbai, 400005, India}

\date{}

\begin{abstract}

We present new observations of two anomalous X-ray pulsars (AXP) \axpa\ and
\axpb\ made in 1998 with the \asca. The energy spectra of these two
AXPs are found to consist of two components, a power-law and a blackbody
emission from the neutron star surface.
These observations, when compared to earlier \asca\
observations in 1994 show remarkable stability in the intensity, spectral
shape and pulse profile. However, we find that the spin-down rate
in \axpb\ is not constant. In this source, we have clearly identified
three epochs with spin-down rates different from each other and the average
value. This has very strong implications for the magnetar hypothesis of AXPs.
We also note that the spin-down rate and its variations in \axpb\ are
much larger than what can normally be produced by an accretion disk with
very low mass accretion rate corresponding to its low X-ray luminosity.

\end{abstract}


\section{Introduction}

Some X-ray pulsars are known to have remarkable similarity in their
properties which are different from other binary or isolated X-ray pulsars
(\cite{mere95b}).
The properties common to most of these objects are
a) pulse period in a small range of 5--12 s,
b) monotonous spin down with ${\rm P} / \dot {\rm P}$ in the range of
$5\times10^{11} - 1.3\times10^{13}$ s
c) identical X-ray spectrum consisting of steep power-law ($\Gamma = 3-4$)
and black body component (kT $\sim 0.5$ keV),  
d) stable X-ray luminosity (10$^{34}-10^{36}$ ergs s$^{-1}$) for years,
e) faint or unidentified optical counterpart,
and f) no evidence of orbital motion.
The sources also have a galactic distribution, most of these are within
$|{b}|\leq 0.5^\circ$
and all are probably young ($\sim 10^4$ yr) because of their association
with SNR or molecular clouds. The objects in which all the properties
mentioned above have been observed are \axpa, 1E~2259+586, \axpb,
1RXS~J170849.0$-$400910 and 1E~1841$-$045 (Kes~73). Two more objects,
AX~J1845.0$-$0300 (\cite{tori98}) and RX~J0720.4$-$3125
(\cite{habe96}),
also probably belong to the same class but to establish their AXP candidacy,
more X-ray observations are required to measure their pulse period variations,
search for possible pulse arrival time delay and investigate the flux
stability. Classification of an object as AXP only from some properties
similar to the above is not very firm. 4U~1626$-$67, probably a binary system,
showed both spin-up and spin-down (\cite{chak97}) and also has an optically
bright accretion disk (\cite{midd81}). Another one object RX~J1838.4$-$0301
(\cite{schw94}), does not have stable intensity and pulsations are also not
always detectable (\cite{song99}). Therefore, these two objects are not AXPs.

Considering the strong similarity between these handful of sources, it
has been proposed that they have same physical nature and different
scenarios have been proposed to explain the observed properties. The
prominent models are
a) accretion from low mass binary companion (\cite{mere95b}),
b) single neutron star accreting from molecular cloud, or a product of
common envelope evolution (Thorne-$\rm \dot{Z}$ytkov object) of close high mass
X-ray binaries in which a solitary neutron star accretes matter from a
fossil disk (van Paradijs, Taam, \& van den Heuvel \cite{vanp95a};
Ghosh, Angelini, \& White \cite{ghos97}), and
c) extremely high magnetic field neutron star radiating X-rays due to magnetic
field decay (\cite{thom96}). Unlike the radio pulsars and rotationally
powered X-ray pulsars, in the AXPs, the spin-down rate is not large enough
to power the observed X-ray emission.

Among the 90 or so known X-ray pulsars (\cite{naga99}), direct evidence of
binary nature is known for more than 35 sources (\cite{vanp95b}). Including
the indirect evidences this number can be upto about 65 and 7 pulsars are
isolated stars in SNR and are powered by rotational energy losses. In the
rest of the pulsars, in which no binary signature is known, it is often due
to lack of sufficient observation. However, the 7 objects which are either
AXPs or candidate AXPs (or 10 if we include the 3 Soft Gamma-ray Repeaters
in which pulsations have been detected), no binary signature has been found
in spite of extensive searches. The strong upper limit on pulse arrival time
delay that has been obtained in some of these sources strongly suggests
non-binary nature for the AXPs. In addition, the spin change behaviour of the
AXPs is also remarkably different from accreting pulsars (\cite{bild97}).
In almost all the accreting X-ray pulsars, both spin-up and spin-down
episodes have been seen which may be randomly distributed (in persistent
sources) or spin-downs in quiescence followed by rapid spin-ups during bright
transient phases (in transient pulsars) or long monotonic spin-up and
spin-down episodes accompanied by spectral and luminosity changes
(e.g. 4U~1626$-$67, \cite{yiiv99}).

The AXPs are in many respect also similar to the X-ray counterparts of the
Soft Gamma-ray Repeaters (SGR). The X-ray spectral and timing properties
of these two type of objects have strong similarities, half of the AXPs and
SGRs are associated with supernova remnants. This has lead to the suggestion
that the AXPs are also magnetars in which the X-ray emission is due to
magnetic field decay (\cite{thom96}). The main difference between
these two type of objects is the non detection of SGR bursts from the AXPs.
However, considering the rarity of the SGR activity among the established
SGRs (\cite{kouv96}), the absence of bursts from AXPs is not a
serious issue. From a relatively young age of the AXP, 1E~1841$-$045 in the
supernova remnant Kes~73, Gotthelf, Vasisht, \& Dotani (\cite{gott99})
proposed that in the evolutionary track, the AXPs are an early quiescent
state of the SGRs. Stability of the X-ray emission properties (spin-down
rate, luminosity, spectral shape and pulse shape and fraction) is usually
mentioned as one important aspect of the AXPs though one has to compare
between observations made with different instruments for which the energy
band, energy resolution and sensitivity are not identical. To make
a rigorous comparison in the stability of the X-ray emission properties
we have made new observations of two AXPs with the \asca, four years
after two previous observations reported by \cite{whit96} and Corbet \&
Mihara (\cite{corb97}). The aim was to critically examine the
stability of the X-ray emission pattern and more pulse period measurements
which may provide support to either the accretion powered or the magnetar
hypothesis for these objects.

The source \axpa\ is close to a long period binary pulsar RX J0146.9+6121
and in the \exo\ observations of this field in which pulsations were first
discovered, pulsations from both the sources were observed simultaneously
(\cite{isra94}). With \exo, the 8.7 s pulsations in this
source were detected only in the 1.0--4.0 keV range.
A binary nature of the system was preferred in spite of a large X-ray to
optical flux ratio and absence of pulse arrival time delay.
The source was subsequently observed with \ros\ and the pulse period
was found to be very close to the \exo\ measurement (\cite{hell94}).
\asca\ observation in 1994 confirmed the steady spin-down trend and
defined the spectral character clearly. A model consisting a 0.4 keV
blackbody and a power law with a photon index of 3.7 was found to
describe the spectrum well. From a
small radius of a few km of the black body emission zone
(which probably is on the
surface of the neutron star) \cite{whit96} suggested that the black-body
component is more likely to be due to a spherical accretion rather than
accretion from a disk. A small
pulse fraction and energy dependent pulse profile, double peak at low energy
and single peak at high energy is characteristic of this pulsar. From \xte\
observations, the upper limit on the pulse arrival time delay was determined
to be 260 ms in the 70 s to 2.5 days range thereby ruling out all types of
binary companions except white dwarf or low mass He main sequence star
(\cite{wils99}). \sax\ observations during 1997--98 confirmed
the spectral characteristics, pulse profiles and spin-down trend
(\cite{isra99b}).  Observations spanning 20 years during 1979--1998, with
the \ein, \exo, \ros, \asca, \sax\ and \xte\ show an overall constant
spin-down trend with $\dot {\rm P}  = 2.2 \times 10^{-12}$ s s$^{-1}$.

Pulsations in the source \axpb\ were discovered from observations
with the \ein\ observatory in 1979 and was confirmed by \exo\ observation
in 1985 (\cite{sewa86}). The energy spectrum was found to be a
power-law type ($\Gamma$ = 2.26) with low energy absorption (N$_{\rm H} = 1.6
\times 10^{22}$ atoms cm$^{-2}$). The relatively harder power-law spectrum and
a candidate optical counterpart lead to the speculation that this can be a Be
star binary. Several \gin\ observations established a secular spin-down trend
with $\dot {\rm P} = 1.5 \times 10^{-11}$ s s$^{-1}$ similar to a few other
soft spectrum pulsars (\cite{corb90}). Subsequent \ros\ observations
however revealed an increase in the spin-down rate which is remarkably
different from other established AXPs (\cite{mere95a}). The low energy part of
the spectrum was first accurately measured with the \asca\ in 1994 (Corbet \&
Mihara \cite{corb97}).
However, it could not distinguish between power-law ($\Gamma =
3.34$) and a combination of power-law ($\Gamma$ = 2.0) and black-body (kT =
0.55 keV). During the \ros\ and the \asca\ observations, the spin-down rate
remained at a higher level of $3.3 \times 10^{-11}$ s s$^{-1}$. A decrease
in intensity by a factor of 3 compared to the \exo\ observation was also
noticed. \sax\ observation in 1997 showed that a combined black-body and
power-law model fits the data well. The size of the black-body emission
zone was found to be of the order of km$^2$, identical to other AXPs
(\cite{oost98}). Compared to other AXPs, the pulse fraction was
found to be much larger ($\sim$ 65\%) in \axpb\ and has little energy
dependence. The power-law component is also relatively harder ($\Gamma = 2
\sim 2.5$) compared to the other AXPs ($\Gamma = 3 \sim 4$). \xte\ observations
in 1996--97 showed yet another change in the spin down rate, now close to that
in 1980's. Long \xte\ observations established a small upper limit of 60 ms for
any pulse arrival time delay for orbital period in the range of 200 s to 1.5
days. Based on this, any binary companion other than low mass helium burning
star in a face on system has been ruled out (\cite{mere98}).

\section{Observations and data analysis}

Both the sources were observed twice with \asca, in 1994 and in 1998 with
about 4 years time difference between the two observations. \asca\ has two
Solid-state Imaging Spectrometers (SIS) and Gas Imaging Spectrometers (GIS)
each at the focal plane of four identical mirrors of typical photon
collecting area 250 cm$^2$ at 6 keV. The energy resolution is 120 eV and
600 eV (FWHM) at 6 keV for the SIS and GIS detectors respectively. For more
details about \asca\ please refer to \cite{tana94}. Details of the 1994
observations are given in Corbet \& Mihara (\cite{corb97}) and
\cite{whit96}. In 1998, the GIS observations were made in normal PH
mode in which the time resolution
is 64 ms and 500 ms at high and medium bit rates respectively. The SIS
observations were made with one of the CCD chips, and has time resolution
of 4 s. The standard data selection criteria of the \asca\ guest observer
facility, that comprises a cut-off rigidity of charged particles 6 GeV/c,
maximum rms deviation from nominal pointing of 0.01 degree, minimum angle
from Earth's limb 10$^\circ$, satellite outside the South Atlantic Anomaly
region etc were applied. Data were removed from the hot and flickering pixels
of the SIS detectors and also the charged particle events were removed
from the GIS detectors based on rise time discrimination. For the two
observations of \axpa, the source photons were extracted from circular
regions of radius 6 arc min and 4 arc min around the source for GIS and SIS
respectively. In the case of \axpb, the source photons were
extracted from relatively smaller regions of 5 and 3 arc min respectively.
For SIS, the background spectra were accumulated from the whole chip
excluding a circular region around the source and for GIS it was collected
from regions diametrically opposite to the source location in the field of
view.

\subsection{Period analysis}

To calculate the pulse periods accurately, barycentric correction was
applied to the arrival time of each photon and light curves were extracted
from the pair of GIS detectors with a time resolution of 0.5 s in the
energy band of 0.5-10.0 keV. Epoch
folding method was applied to obtain the pulse periods approximately and
templates for the pulse profiles were created by folding the light curves
at the approximate pulse periods. Subsequently, the light curves were
divided into eight segments of equal length and pulse profiles were created
from each of these segments by applying the same epoch and pulsation period.
The relative phases of the pulses were then
evaluated by cross correlating the pulse profiles with the respective
templates. A linear fit to the relative phases with their pulse numbers gave
the correction necessary to obtain the accurate pulse period. The 1998
observations of the two AXPs resulted in new measurement of pulse periods
at these epochs and from the 1994 observations we obtained similar pulse
periods as reported earlier, with reduced uncertainty for \axpa. The
pulse periods obtained for the two sources are given in Table 1. The pulse
profiles of the two sources in three energy bands are shown in Figure 1
and Figure 2. The pulse fraction, defined as the ratio of the pulsed to total
flux, was calculated from background subtracted pulse profiles in the 0.5--10.0
keV band. Pulse fractions were found to be identical in both the observations,
$\sim 9\%$ and $\sim 75\%$ in \axpa\ and \axpb\ respectively. In \axpa, the
pulse profile shows energy dependence, from double peaked in low
(0.5--1.5 keV) and medium (1.5--4.0 keV) energy to single peaked in high
(4.0--8.0 keV) energy, associated with increase in the pulse fraction.
In \axpb, on the other hand, the pulse profile and pulse fraction are almost
identical over the \asca\ energy range. Light curves of the two sources
did not show any intensity variations at minutes to days time scale.

\begin{figure}[t]
\centerline{\psfig{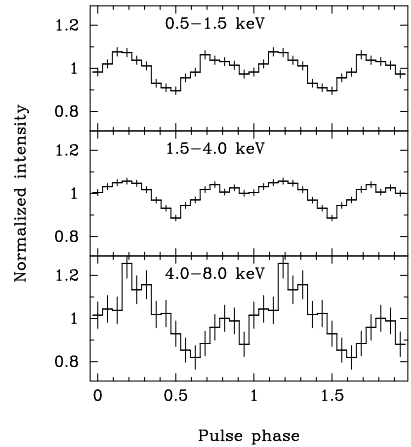}}
\caption
{The background subtracted pulse profiles of \axpa\ obtained
with the GIS in three energy bands are plotted for two cycles. A change
in the pulse profile, from double peaked at low energies to single peaked
at high energies can be noticed.}\label{fig1}
\end{figure}

\begin{figure}[t]
\centerline{\psfig{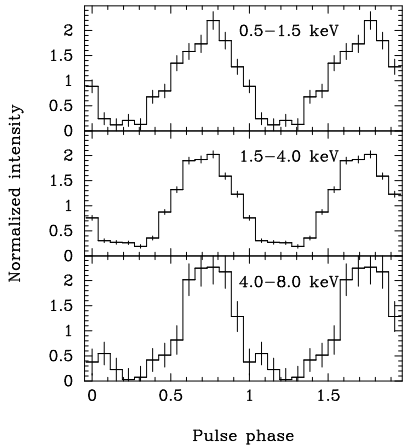}}
\caption
{The background subtracted pulse profiles of \axpb, similar
to Figure 1. The pulse fraction of this source is $\sim$75\%, highest among
the AXPs.}\label{fig2}
\end{figure}

\subsection{Spectral analysis}

\begin{figure}[t]
\centerline{\psfig{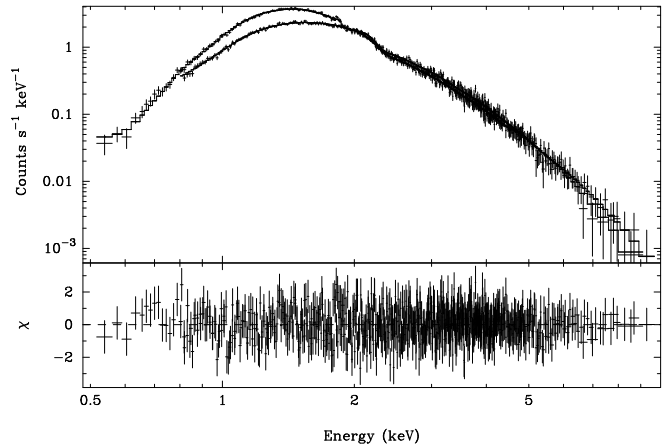}}
\caption
{The observed SIS and GIS energy spectra of \axpa\ shown with
histograms for the model spectra folded through the responses matrices
and the residuals.}\label{fig3}
\end{figure}

\begin{figure}[t]
\centerline{\psfig{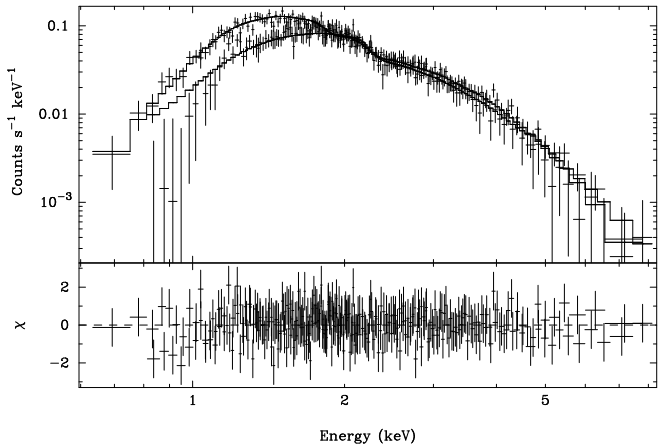}}
\caption
{The SIS and GIS energy spectra of \axpb\ along with the
residuals. The histograms represent the best fitted model folded with the
response matrices. A simultaneous fitting of the SIS and GIS spectra was
performed.}\label{fig4}
\end{figure}

In \axpa, a simple absorbed power-law fit shows large residuals at low
energy and addition of a black-body component results in acceptable fit.
For the 1998 observation, inclusion of the black-body component improved
the reduced $\chi ^2$ from  2.88 to 0.94 and 1.23 to 0.76 for the GIS and SIS
respectively. Similar improvement in fitting was reported by \cite{whit96}
for the 1994 observation.
The photon index of the power-law component obtained from the SIS data
for the two observations are identical, $\Gamma = 3.3$. The photon index
from the GIS data is slightly different from the SIS value, $\Gamma = 3.9$,
but it is identical for the two observations. The temperature of the black
body component obtained from both SIS and GIS are identical, 0.39 keV,
in both the observations and the flux in the two components are also
identical. The difference in the photon index between the SIS and GIS can
possibly be attributed to calibration uncertainty, because the in-flight
spectral calibration of the spectrometers are done with sources which have
relatively harder spectra. The SIS and GIS spectra of the 1998 observation
are shown in Figure 3 along with the best fitted models for the respective
detectors and the residuals to the model spectra.

For the other source \axpb, Corbet and Mihara (\cite{corb97}) showed that
the power-law and black-body with power-law models could not be
distinguished from spectral analysis of the \asca\ data. We have found
different and deeper minima in the $\chi ^2$ for the black-body with
power-law model. The improvement in $\chi ^2$, ($\Delta\chi^2$ of 41 for
292 degrees of freedom and 28 for 261 degrees of freedom for the 1998 and
1994 observations respectively, both indicating probability of chance
occurrence less than 10$^{-3}$) therefore favours the inclusion of a
black-body component. The black-body component has also been detected with
\sax\ (\cite{oost98}). The best fitted power-law plus black-body model
gives a photon index of $\sim 3.0$, black body temperature of $\sim 0.56$
keV and column density of $\sim 1 \times 10^{22}$ atoms cm$^{-1}$. The
spectral parameters are identical in both the \asca\ observations and are
given in Table 2. Due to relative weakness of this source, 
simultaneous spectral fitting was carried out with the GIS and SIS data
for the 1998 observation (Figure 4). However, for the 1994 observation,
the SIS were operated in FAST mode and spectra were available only
from the two GIS detectors.

\section{Discussion}

\subsection{Period changes and Emission stability}

The pulse period measurements of \axpa\ is rather scarce except for the
last two years (Figure 5). In spite of the source being very bright, with a
flux of more than 10$^{-10}$ erg cm$^{-2}$ in the 2.0--10.0 keV band, a pulse
fraction of only about 10\% has restricted the individual pulse period
measurements accurate to only about ${{\Delta {\rm P}} / {\rm P}} \sim 
$ 10$^{-5}$. A linear fit to the pulse period history
shows that of the 10 measurements available, only during the 1994 \asca\
observation the pulse period measurement was slightly different (2.5$\sigma$)
from the linear trend. If the reported errors of all the measurements are
taken at their face value, the linear fit gives a reduced $\chi^2$ of 1.7
for 8 degrees of freedom. It therefore can be concluded that the recent \asca\
measurement together with the previous results is consistent with a
constant spin-down rate. The observations are not yet sufficient to clearly
identify any significant variation from a constant spin-down rate.

In the source \axpb, departure from a linear spin-down is already
known (\cite{mere95a}; \cite{oost98}). In this source, an order of
magnitude larger spin-down rate and better pulse period measurements
$({{\Delta {\rm P}} / {\rm P}} \sim $ 10$^{-6})$, owing to a high pulse
fraction $(\sim 75\%)$, help us to identify three different epochs of
spin-down history. Though the method adopted for calculating the errors
in the pulse period is not known for all the observations and the
uncertainty level is likely to be nonuniform, a constant spin-down
trend can be ruled out without any doubt. A linear fit to the pulse
period history with the reported errors gives a reduced $\chi ^2$ of
4500 for 11 degrees of freedom. Including the 1998 \asca\ observation
(see Figure 6) with the recent \sax\ and \xte\ observations, we
find that from 1996 the source has a spin-down rate of $(1.67 \pm 0.02)
\times 10^{-11}$ s s$^{-1}$. This is a factor of 2 smaller than the
spin-down rate of $(3.29 \pm 0.03) \times 10^{-11}$ s s$^{-1}$ during
the 1994--1996 period. The present spin-down rate is closer to the value
of $(1.5 \pm 0.5) \times 10^{-11}$ s s$^{-1}$, measured during the \ein,
\exo\ and \gin\ observations made in the period 1979--1988. The spin-down
rate is much closer to being constant during these three epochs with
reduced $\chi ^2$ of 0.8, 7.7, and 36 for 3, 3, and 1 degrees of freedom
respectively.

These two sources do not show flux variability on time scales from
a few minutes to days. In the \asca\ observations of both the sources
separated by 4 years we have found that the overall intensity and spectral
parameters have remarkable stability. A difference between the GIS and SIS
photon index that has been found in \axpa, is due to calibration
uncertainties. The spectral parameters obtained from the 1998 GIS and
SIS observations are identical to the 1994 values. The spectral parameters
obtained from the simultaneous fitting of the GIS and SIS spectra are
similar to the \sax\ values obtained during 1997--1998.
In \axpa, the flux history shows a rms variation of 15\% around
the average value (Figure 5), and multiple measurements with the same
instrument (\asca\ and \sax) gave almost identical flux.
In \axpb, the over all intensity during the two \asca\ observations
and one \sax\ observation in between are within $10\%$ of the average value.
The 2.0--10.0 keV fluxes during the \ein\ and \ros\ observations
are estimated by extrapolating the measurements in the low energy
bands of 0.2--4.0 keV and 0.5--2.5 keV respectively, and using a
rather low photon index of 2.26, obtained by \exo.
The flux during the \gin\ observation is estimated by comparing the
pulsed fluxes during the \gin\ and \exo\ observations and assuming
that the pulse fraction remained same.
The flux measurements from the previous observations as shown in
the bottom panel of Figure 6 are about a factor 3 higher than the recent
measurements with \asca\ and \sax.
We note that there is some overestimation in extrapolation of the soft
X-ray measurements with \ein\ and \ros\ due to a smaller photon index used,
and the flux estimates from \exo\ and \gin\ could be overestimated due
to contribution from the nearby bright and variable source $\eta$-Carina,
which is only 0.4$^\circ$ away and is about 15 times brighter than \axpb\
(\cite{corc98}; \cite{ishi99}).
In view of the stability of the flux during
1994--1998, as obtained from the imaging instruments \asca\ and
\sax, it is possible that the overall intensity of this source
does not vary at a few years times scale. The spectral parameters from
the two \asca\ observations are also identical and consistent with the values
obtained with \sax\ in between. However, we note that in \axpb,
the absorption column density obtained from the two \asca\ observations are
identical and a factor of 1.5--2 larger than the \sax\ measurement in
between. This difference is somewhat larger than the known calibration
difference between the two instruments (\cite{orra98}).

Some Low Mass X-ray Binaries (LMXB) have also been detected at low
luminosity levels ($\sim10^{33}$ erg s$^{-1}$; e.g, Cen~X$-$4, Aql~X$-$1,
see \cite{tana96} and references therein), understood to be quiescent phase
of the Soft X-ray Transients (SXT). Usually the LMXB sources show both short
and long term irregular intensity variations and many also show quasi-periodic
oscillations, bursts, dips or orbital modulation. Even though the AXPs
do not show significant temporal variations other than the pulsations, which
is somewhat different from typical characteristics of low-luminosity LMXBs,
this itself is not a strong argument against the AXPs being LMXBs.    
One LMXB which has some properties similar to the AXPs is 4U 1626$-$67. The
X-ray luminosity, magnitude of spin-change rate, pulse period, and flux
stability over very short to years time scale of this source are similar
to the AXPs. But, presence of both spin-up and spin-down, quasi-periodic
oscillations, optically bright accretion disk, and a hard X-ray spectrum
makes it different from the AXPs.

\begin{figure}[t]
\centerline{\psfig{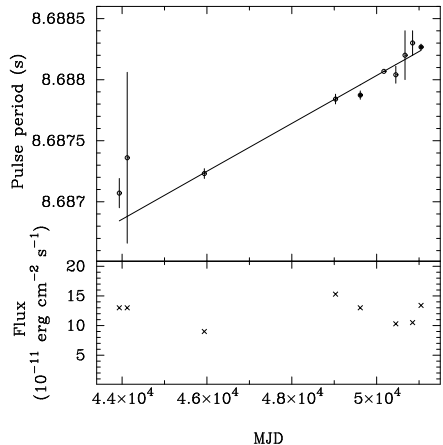}}
\caption
{The pulse period and flux history of \axpa. The straight line
shows the best fit for a constant spin-down. The pulse period
measurements with {\it ASCA} are marked with filled circles and
the open circles are for all the other observations mentioned in the
text.\label{fig5}}
\end{figure}

\begin{figure}[t]
\centerline{\psfig{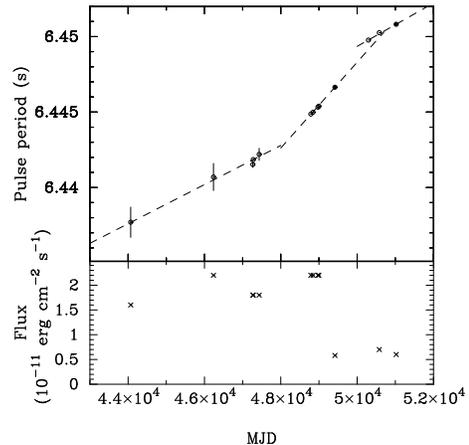}}
\caption
{The pulse period and flux history of \axpb. The lines are
used to identify the different spin-down epochs. The pulse period
measurements with {\it ASCA} are marked with filled circles
and other observations with open circles.}\label{fig6}
\end{figure}

\begin{figure}[t]
\centerline{\psfig{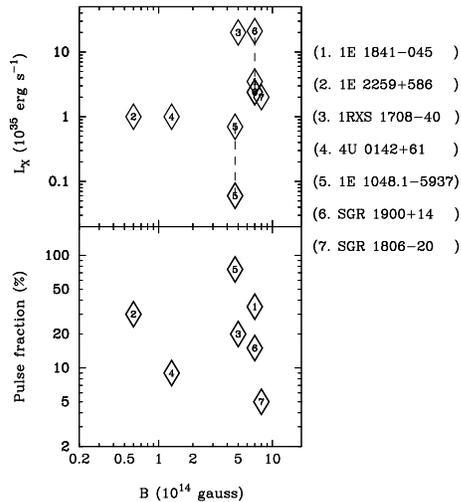}}
\caption
{The X-ray luminosity and pulse fraction of the AXPs and SGRs are
plotted against the magnetic field strength assuming that these objects
are magnetars. Two values of the luminosity are plotted for \axpb\
to show the uncertainty in its distance and same has been done for \sgrb\
to show the variability in its X-ray emission. Distance of
1RXS~J170849.0$-$400910 is taken to be 10 kpc and that for other sources are the
best available estimates.}\label{fig7}
\end{figure}

\subsection{Accretion torque in the common envelope evolution model}

It has been proposed that the AXPs are recent remnants of common-envelope
evolution of high-mass X-ray binaries (van Paradijs et al. \cite{vanp95a};
Ghosh et
al. \cite{ghos97}). In this model, the pulsar is rotating near its equilibrium
period, close to the Keplerian period at the innermost part of the disk.
If the pulsar is rotating at the equilibrium period, small changes in the
mass accretion rate cause alternate spin-up and spin-down episodes. The
over all spin-down is explained with the assumption that in the absence of
a companion star as the source of mass accretion, the mass accretion rate
from the disk decreases slowly on viscous time scale. The equilibrium period
of the pulsar is inversely related to the mass accretion rate and shows
secular increase. For \axpb, which has a pulse period of 6.5 s and
luminosity of $6.3 \times 10^{33}$ erg s$^{-1}$ for a distance of 3 kpc
(or in a more favourable case, $7 \times 10^{34}$ erg s$^{-1}$ if the
source is at a distance of 10 kpc), the magnetic field strength inferred
for an equilibrium rotator (\cite{fran92}) is
${\rm B} = 10^{11}~({\rm P / 3})^{7 / 6}~{\rm L_{35}}^{1 / 2} = 
6.2 \times 10^{10}$ (or $2.1 \times 10^{11}$) gauss. Here, and in what
follows we have assumed a neutron star with mass M = 1.4 ${\rm M}_{\sun}$,
radius R = $10^6$ cm, and moment of inertia I = $10^{45}$ gm cm$^2$. A
pulse period of P implies that if in equilibrium, the co-rotation radius or in
this case the radius of the inner disk is ${\rm r_M} = ({\rm GM})^{1 / 3}~
({{\rm P} / {2\pi}})^{2 / 3}$. The infalling material from the disk
carries a positive angular momentum of $\dot {\rm M}~({\rm GMr_M})^{1 / 2}$.
The torque can also be expressed in terms of the pulse period P and the
X-ray luminosity ${\rm L}_X$ as ${\rm RL_X~{({\rm P / {2\pi GM}})}^{1
/ 3}}$. Even if we assume that all of the X-ray emission is a result of
disk accretion, the accretion torque is only 1.1 $\times 10^{31}$
(or 1.2 $\times 10^{32}$) gm cm$^2$ s$^{-2}$. To achieve the
observed spin-down rate for a neutron star with moment of inertia $10^{45}$
gm cm$^2$, the negative torque required to be imparted onto the neutron star
is I$\dot \Omega = 4.9 \times 10^{33}$ gm cm$^2$ s$^{-2}$. This is a factor
of 450 (or 40) larger than the accretion torque, and in the common envelope
evolution model, a negative dimensionless torque of this magnitude is required
to spin-down the pulsar at the observed rate. In other words, the spin-down
rate of this source is much larger than what can be achieved with disk
accretion onto a neutron star with a luminosity of less than $10^{35}$ erg
s$^{-1}$. \axpb\ does not fit in the classical picture of $\dot
{\rm P}~vs~{\rm PL}^{3 / 7}$ of the equilibrium rotators (\cite{ghos79}).
In addition, the two \asca\ observations made during the two epochs which have
a factor of 2 different spin-down rates, do not show significant difference
in the luminosity or the spectral parameters.

Among the other AXPs, 1RXS~J170849.0$-$400910 (\cite{sugi97}; \cite{isra99a})
and 1E~1841$-$045 (Kes~73, \cite{vasi97}; Gotthelf et al. \cite{gott99}) have relatively large spin-down rate of $2.2 \times
10^{-11}$ and $4.1\times 10^{-11}$ s s$^{-1}$ respective, while the X-ray
luminosity is in the range of $10^{35} - 10^{36}$ erg s$^{-1}$. In the common
envelope evolution model of the AXPs, a faster spin-down compared to the
torque provided by the accreting matter can be a potential problem for these
two sources also. For this model to be correct for the AXPs, a very fine
tuning of the mass accretion rate is required. All the sources need to have
disk accretion rate a tiny fraction larger than the propeller regime.
\cite{lixd99} have identified several other problems in the context of
applicability of this model to the AXPs. Most notable is the limited lifetime
of an accretion disk around a solitary neutron star compared to the response
time scale of the neutron star to changes in the accretion torque. One possible
alternative is that the spin-down is due to magnetically driven wind from
an accretion disk, proposed also for \sgrb\ (\cite{mars99}), but this will
require a harder X-ray spectrum than what has been observed. \cite{chat99}
proposed a scenario in which the AXPs are formed as single neutron stars
with fossil disks made from fallback material from the supernovae explosions.
In this model, the spin-down from an initial period of a few ms to $\sim$6 s
is due to strong propeller effect at some period when the accretion rate is
very low. But, if accretion is the correct phenomenon in the AXPs, as high
spin-down goes on in presence of substantial accretion, spin-down due to
wind outflow seems to be more plausible than accretion induced angular
momentum loss. However, it should be remembered that the classical equilibrium
disk picture assumed here is often found not to be the most appropriate
description for the X-ray pulsars (\cite{bild97}).

\subsection{Magnetar model}

In view of a very narrow mass and type allowed for any binary companion,
and several arguments against the common envelope evolution model, the magnetar
model seems to be the most likely one for the AXPs. If the spin-down is
due to magnetic braking, dipole field strength of the order of $10^{14}$
gauss is estimated for these sources. A nearly constant spin-down property
was thought to favour the magnetar model over a binary scenario. In two
AXPs, 1E~1841$-$045 (Gotthelf et al. \cite{gott99}), and 1RXS~J170849.0$-$400910
(Kaspi, Chakrabarty \& Steinberger \cite{kasp99}), there is very strong
evidence for constant spin-down, whereas in \axpb, deviation from a linear
trend is clear. Recently, a deviation has also been detected from \sgrb\
(Woods et al. \cite{wood99}). 1E~2259+586, the most frequently observed AXP,
has provided an interesting pulse period history. Observations made for
about 15 years with many instruments preceding \xte\ showed considerable
variation in the spin-down rate (\cite{bayk96}). But, the pulse-coherent
timing observations with \xte\ proved it to be otherwise, at least for a
period of last 2.6 yr (Kaspi et al. \cite{kasp99}).

Two scenarios have been proposed which can explain the changing spin-down
rate even when the overall braking is due to the ultrastrong magnetic field.
\cite{mela99} showed that for reasonable neutron star parameters, a
radiative precession effect may take place which
can give the observed spin-down variations with time scale of about 10 years.
It will be possible to verify this scenario when more pulse period
measurements become available in the next few years. Woods et al.
(\cite{wood99}) have 
identified a possible {\it braking glitch} in \sgrb\ close to the time
when SGR activity was very strong. But if the spin-down variation is
related to the SGR activity, similar activities should have been observed
from \axpb\ and 1E 2259+586. We have found that all the gamma ray
bursts observed with the BATSE for which the estimated positions are
within $2\sigma$ of these AXPs (about 30 GRBs around each AXP),
have strong high energy emission unlike the
SGR bursts. \cite{heyl99} have proposed
that the spin-down variations can be explained as glitches (similar to radio
pulsars) superposed on constant spin-down. But, with recent pulse period
measurements of \axpb\ and \sgrb, this will require too many
glitches, one before almost every observation unless there are {\it braking
glitches}, never observed in radio pulsars. 

In the magnetar model, the X-ray emission is due to decay of the magnetic
field. The energy generated at the core is transported to the crust along
the magnetic field direction. The black body component of the spectrum is
thermal emission from the hot spots at the magnetic polar regions and the
power-law component is part of the thermal emission reprocessed by the
magnetic field and the environment. Investigation is required about the
expected pulse profile and its energy dependence. Time and/or energy
dependence of the pulse profile as has been observed in \sgra\
(\cite{kouv98}) and \sgrb\ (\cite{hurl99};
\cite{kouv99}; \cite{mura99}) also requires to be addressed. A double
peaked pulse profile at low energy and single peaked profile at high energy
are observed in \axpa\ (Figure 1). The pulsation is very weak in some
sources (only 5--10$\%$ in \sgra\ and \axpa), and 75$\%$
in \axpb. If the pulsation is due to confinement of the
heat in the magnetic polar regions by the magnetic field, a correlation
between magnetic field strength and pulse fraction should be observed.
But, it is likely to be smeared by the geometric effect of individual
sources, i.e. the orientation of the spin and magnetic axes with respect to
the line of sight. In the magnetar model, there are two mechanisms by which
X-rays can be generated. If the X-ray emission is powered by decaying
magnetic field, the luminosity is a very strong function of the magnetic
field strength, ${\rm L_x \propto B^4}$ (\cite{thom96}). Alternate
process of X-ray generation is particle acceleration by Alfven waves
resulting from small scale fracture of the crust, in this case ${\rm L_x
\propto B^2}$. The X-ray luminosity and pulse fraction of five confirmed
AXPs and two SGR sources are shown in Figure 7. against the magnetic
field strength. The later is estimated from pulse period and the overall
spin-down rate assuming ${\rm B = 3.2 \times 10^{19}~(P\dot P)^{1 / 2}}$
gauss. Though there is uncertainty in the luminosity of some
sources (see the caption of Figure 7), a 2nd or 4th power correlation between
${\rm L_X}$ and B does not seem to be present. There is also no correlation
between pulse fraction and the magnetic field.

A clustering of the pulse period of 10 sources (7 AXPs and 3 SGRs) in the
5--12 s range also needs to be addressed, when the magnetars are expected
to be alive in X-ray until they have slowed down to a pulse period of about
70 s (\cite{dunc92}). The magnetars are expected to be radio quiet
due to suppression of pair creation at high magnetic field (\cite{bari98}),
and this seems to be true for most sources.

Changes in luminosity or small changes in the spectral parameters can be a
result of varying activities in the core, the heat generated from which is
transported to the surface along the direction of the magnetic field at time
scale of a few years. Variability study of AXPs with very high sensitivity
may rule out magnetar model if significant change in column density is
observed. This will indicate the presence of accretion disk and wind in the
neighbouring area. The \sax\ and \asca\ observations give identical column
density for \axpa\ but slightly different column density in case of
\axpb. However, multiple observation of the later source with the
same instrument is found to give identical value indicating that the
difference between the two instruments can also be a systematic effect.

\section{Conclusion}

Using multiple observations of two AXPs with the {\it \asca} we have found
remarkable stability in the intensity, spectral shape and pulse profile
over a 4 years period. For the source \axpb, we have
confirmed that similar to other AXPs, the spectrum consists of a power-law
component and a black body component. The spin-down trend of \axpa\ is
consistent with a constant rate whereas in \axpb\ we have clearly
identified three different spin-down epochs. We have shown that the fast
spin-down of some of the AXPs is difficult to achieve with disk accretion.
Hence the common envelope evolution scenario of AXPs is unlikely to be the case,
unless the spin-down is due to wind outflow from the disk. In this case
also, the stability of X-ray emission in spite of varying spin-down rate
remains unexplained. The significantly different spin-down rates of
\axpb\ in different epochs are also difficult to reconcile in the
magnetar model, unless precession is at work. But the flux stability and
lack of variation of the spectral parameters appear to favour the magnetar
model.

\begin{acknowledgements}

We thank an anonymous referee for many suggestions which helped to improve
a previous version of the manuscript. B. Paul was supported
by the Japan Society for the Promotion of Science through a fellowship.

\end{acknowledgements}

{}

\clearpage

\begin{deluxetable}{ccc}
\footnotesize
\tablenum{1}
\tablecaption{The pulse periods from the \asca\ observations\label{tbl-1}}
\tablewidth{0pt}
\tablehead{
\colhead{Date of Observation}&\colhead{Source name}&\colhead{Pulse period}\nl
\colhead{(MJD)}&\colhead{}&\colhead{(s)}\nl}
\startdata
		&	{\bf \axpa}&\nl
49614.1		&&	8.687873 $\pm$ 0.000034	\nl
51046.7		&&	8.688267 $\pm$ 0.000024	\nl
\nl
\hline
\nl
		&	{\bf \axpb}&\nl
49416.5		&&	6.446645 $\pm$  0.000001\nl
51021.1		&&	6.450815 $\pm$  0.000002\nl
\nl

\enddata

\end{deluxetable}

\begin{deluxetable}{lccccccc}
\footnotesize
\tablenum{2}
\tablecaption{The spectral parameters\label{tbl-2}}
\tablewidth{0pt}
\tablehead{
\colhead{}&\colhead{}&\colhead{}&\colhead{\bf \axpa}&\colhead{}&\colhead{}&\colhead{}\nl
\colhead{Obs. date}&\colhead{}&\colhead{1994/09/18-19}&\colhead{}&\colhead{}&\colhead{1998/08/21}&\colhead{}\nl
\colhead{}&\colhead{GIS}&\colhead{SIS}&\colhead{SIS+GIS}&\colhead{GIS}&\colhead{SIS}&\colhead{SIS+GIS}\nl
}

\startdata
N$_{\rm H}$\tablenotemark{a}       & 1.03 $\pm 0.08$ & 0.92 $\pm 0.05$ & 1.10 $\pm 0.04$ & 1.08 $\pm 0.09$ & 0.97 $\pm 0.08$ & 1.17 $\pm 0.04$ \nl
Photon index      & 3.9 $\pm 0.1$ & 3.3 $\pm 0.1$ & 3.84$\pm 0.08$ & 3.98 $\pm 0.15$ & 3.3 $\pm 0.2$ & 3.87 $\pm 0.09$ \nl
Power-law norm\tablenotemark{b}    & 0.26 $\pm 0.05$ & 0.12 $\pm 0.03$ & 0.25 $\pm 0.03$ & 0.30 $\pm 0.06$ & 0.10 $\pm 0.03$ & 0.24 $\pm 0.03$ \nl
BB temp (keV)              & 0.39 $\pm 0.01$ & 0.380 $\pm 0.006$ & 0.382 $\pm 0.007$ & 0.399 $\pm 0.014$ & 0.384 $\pm 0.005$ & 0.378 $\pm 0.009$ \nl
BB norm\tablenotemark{c}~~~~~~~~~~~ & 1.4 $\pm 0.2$ & 1.9 $\pm 0.1$ & 1.5 $\pm 0.1$ & 1.16 $\pm 0.18$ & 1.97 $\pm 0.15$ & 1.30 $\pm 0.12$ \nl 
Reduced $\chi ^2$ / dof & 0.95 / 691  & 1.11  / 356 & 1.35 / 1030 & 0.76 / 354 & 0.94 / 287 & 1.76 /645 \nl
Observed flux\tablenotemark{d} && 13.0             &&& 13.4 \nl

\nl
\hline
\nl

&&&{\bf \axpb} \nl
Obs. date&&1994/03/02-05&&&1998/07/26-27 \nl
&GIS&&&&&SIS+GIS \nl
\nl
\hline
N$_{\rm H}$\tablenotemark{a}      & 1.0 $\pm 0.2$    &&&&& 1.21 $ \pm0.24$ \nl
Photon index                      & 2.9 $\pm 0.3$    &&&&& 3.2 $\pm 0.5$ \nl
Power-law norm\tablenotemark{b}  & 5 $\pm 3~10^{-3}$&&&&& 7 $\pm 4~10^{-3}$ \nl
BB temp (keV)             & 0.57 $\pm 0.04$  &&&&& 0.56 $\pm 0.06$ \nl
BB norm\tablenotemark{c} & 0.05 $\pm 0.02$    &&&&& 0.060 $\pm 0.015$ \nl
Reduced $\chi ^2$ / dof           & 0.98 /261 &&&&& 0.62 / 292 \nl
Unabsorbed flux\tablenotemark{d} & 0.58              &&&&& 0.60 \nl

\tablenotetext{a}{$10^{22}$ atoms cm$^{-2}$}
\tablenotetext{b}{photons cm$^{-2}$ s$^{-1}$ keV$^{-1}$ at 1 keV}
\tablenotetext{c}{$10^{-3}$ photons cm$^{-2}$ s$^{-1}$}
\tablenotetext{d}{$10^{-11}$ ergs cm$^{-2}$ s$^{-1}$, 2-10 keV}
\enddata

\end{deluxetable}

\end{document}